\documentclass[12pt,a4]{article}
\pdfoutput=1
\usepackage[textheight=.76\paperheight, textwidth=.76\paperwidth, a4paper]{geometry}
\usepackage[bf,small,raggedright]{titlesec}
 \csname
@addtoreset\endcsname{equation}{section}

\usepackage[utf8]{inputenc}
\usepackage{latexsym, graphicx} 
\usepackage{amsmath,amsfonts,amssymb}
\usepackage{caption,subcaption}
\usepackage{cite}
\usepackage[colorlinks=true,linkcolor=blue,citecolor=blue,urlcolor=blue,filecolor=black]{hyperref}

\newcommand{\Geff}{G_{\text{eff}}}
\newcommand{\Heff}{H_{\text{eff}}}
\newcommand{\Muv}{M_{\text{UV}}}

\begin{document}

\begin{titlepage}
\unitlength = 1mm~\\
\vskip 3cm
\begin{center}

{\large{\textsc{Robustness of the Cosmological Constant Damping Mechanism Through Matter Eras}}}

\vspace{0.8cm}
Oleg Evnin{$^{a,b}$}, Victor Massart{$^{c}$}, Kévin Nguyen{$^{d,e}$}\\
\vspace{1cm}

{$^a$\it Department of Physics, Faculty of Science, Chulalongkorn University, Bangkok, Thailand\\}
\vskip 2mm
{$^b$\it Theoretische Natuurkunde, Vrije Universiteit Brussel and\\
The International Solvay Institutes, Brussels, Belgium\\}
\vskip 2mm
{$^c$\it Groupe de physique des particules, D\'epartement de physique\\ Universit\'{e} de Montr\'{e}al, Montr\'{e}al Qu\'ebec, Canada\\}
\vskip 2mm
{$^d$\it Black Hole Initiative, Harvard University, Cambridge, USA} \\
\vskip 2mm
{$^e$\it Department of Mathematics, King's College London, London, United Kingdom} \\
\vspace{0.5cm}
{email:\href{mailto: oleg.evnin@gmail.com}{\hskip 2mm oleg.evnin@gmail.com},\href{mailto: massart.victor@live.be}{\hskip 2mm massart.victor@live.be},\href{mailto: kevin.nguyen@kcl.ac.uk}{\hskip 2mm kevin.nguyen@kcl.ac.uk}}

\vspace{0.8cm}

\begin{abstract}
A dynamical resolution to the cosmological constant fine-tuning problem has been previously put forward, based on a gravitational scalar-tensor theory possessing de Sitter attractor solutions characterized by a small Hubble expansion rate, irrespective of an initially large vacuum energy. We show that a technically natural subregion of the parameter space yields a cosmological evolution through radiation- and matter-dominated eras that is essentially indistinguishable from that predicted by General Relativity. Similarly, the proposed model automatically satisfies the observational constraints on a fifth force mediated by the new scalar degree of freedom.   
\end{abstract}

\end{center}

\end{titlepage}
  
\section{Introduction}
The naturalness problem associated with the cosmological constant $\Lambda$ remains one of the great puzzles faced by theoretical physics. The huge discrepancy between theoretical estimates and actual observations also places this puzzle among the most striking ones. The origin of the problem lies in the gravitational effects of the vacuum energy associated with quantum fields when considered in conjunction with General Relativity (GR). Indeed, the framework of Quantum Field Theory (QFT) in curved spacetime predicts that any such vacuum energy precisely behaves as the cosmological constant introduced long ago by Einstein, and entering the standard $\Lambda$CDM model of cosmology as a free parameter; see the excellent review \cite{Martin:2012bt}. Assuming that General Relativity fully describes the dynamics of spacetime on cosmological scales, one is led to the conclusion that the late-time value of the Hubble expansion rate $H(t)$ observed in the Universe is linked directly to the value of the cosmological constant $\Lambda$,
\begin{equation}
H_0^2\equiv \lim_{t\to \infty} H(t)^2=\frac{\Lambda}{3}.
\end{equation}
Hence, the value of the vacuum energy density $\rho_\Lambda$ may be inferred from measurements of the Hubble expansion rate, yielding \cite{Aghanim:2018eyx}
\begin{equation}
\label{eq:rho-observed}
\rho_{\Lambda}\equiv \frac{\Lambda}{8\pi G}\sim 10^{-47}\, \text{GeV}^4,
\end{equation}
where $G$ is Newton's constant. Such a small value appears \textit{unnatural} from the low-energy standpoint of Effective Field Theory (EFT). In this framework, one expects physics beyond the Standard Model to appear at some ultra-violet (UV) energy scale $\Muv$, and the predictive power of the low-energy theory at hand should not depend on fine details of the theory at this much higher UV scale. Thus, one relies on a decoupling of energy scales in the study of physical phenomena. A simple way to estimate the dependence of any observable quantity on the physics at the scale $\Muv$ is to compute radiative corrections with explicit UV cutoff given by $\Muv$. In particular, any massive particle with mass $m$ gives a loop correction to the vacuum energy density, of the form
\begin{equation}
\delta \rho_{\Lambda} \sim c_1\, \Muv^4 + c_2\, m^2 \Muv^2 +c_3\, m^2 \ln \frac{m}{\Muv}+...\, ,
\end{equation}
where $c_i$ are $O(1)$ constants. Since the scale of new physics is necessarily above the electroweak scale, the above loop correction is at least of order
\begin{equation}
\label{eq:delta-rho}
\delta \rho_\Lambda \gtrsim 10^8\, \text{GeV}^4.
\end{equation}
This quantum contribution appears much larger than the observed value \eqref{eq:rho-observed}, and it looks like only an incredible amount of fine-tuning in the UV-complete theory -- if such a thing even exists -- can explain this observed small value. Although one cannot rule out the possibility of fine-tuning entirely, it shows at least that the sensitivity of the cosmological constant to the physics at high energies is extremely strong. This has motivated the search for cosmological models achieving a dynamical \textit{self-tuning} of the Hubble expansion rate to the small observed value, even in the presence of a large vacuum energy density; for some recent work in this direction, see \cite{Alberte:2016izw,Appleby:2018yci,Emond:2018fvv,Appleby:2020njl,Brax:2019fgj,Lombriser:2019jia,Sobral-Blanco:2020rdu,Babichev:2016kdt}. We also refer the reader to the review \cite{Hebecker:2020aqr} for a discussion of the cosmological constant problem from the EFT viewpoint, and to the literature on effective field theories of dark energy and quintessence \cite{Park:2010cw,Bloomfield:2011np,Gubitosi:2012hu,Bloomfield:2012ff,Gleyzes:2013ooa,Kennedy:2019nie,Renevey:2020tvr} for a complementary perspective.

\begin{figure}
\centering
	\begin{subfigure}[b]{.49\textwidth}
		\centering
		\includegraphics[trim=8cm 8cm 8cm 8cm, scale=0.6]{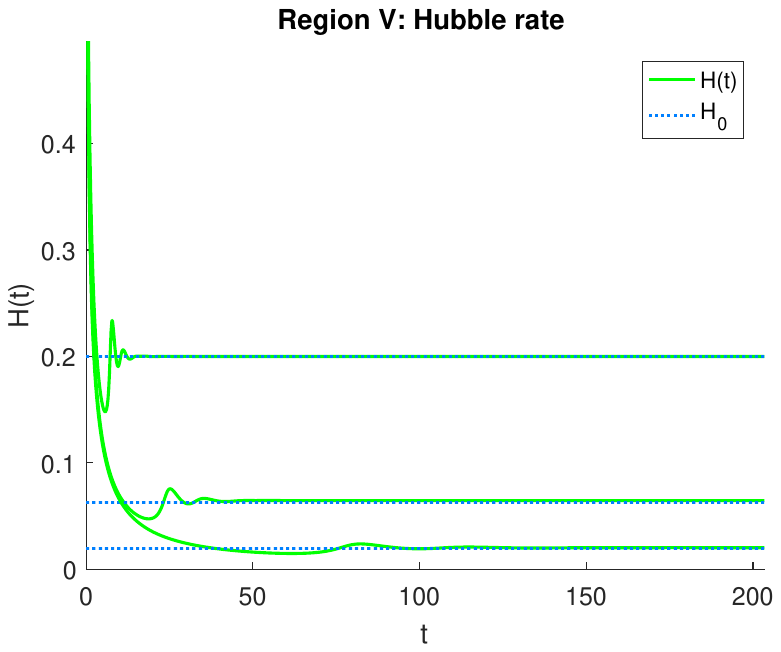}
		\caption{}
		\label{fig:Region5Hubble}
	\end{subfigure}
	\begin{subfigure}[b]{.49\textwidth}
		\centering
		\includegraphics[trim=8cm 8cm 8cm 8cm, scale=0.6]{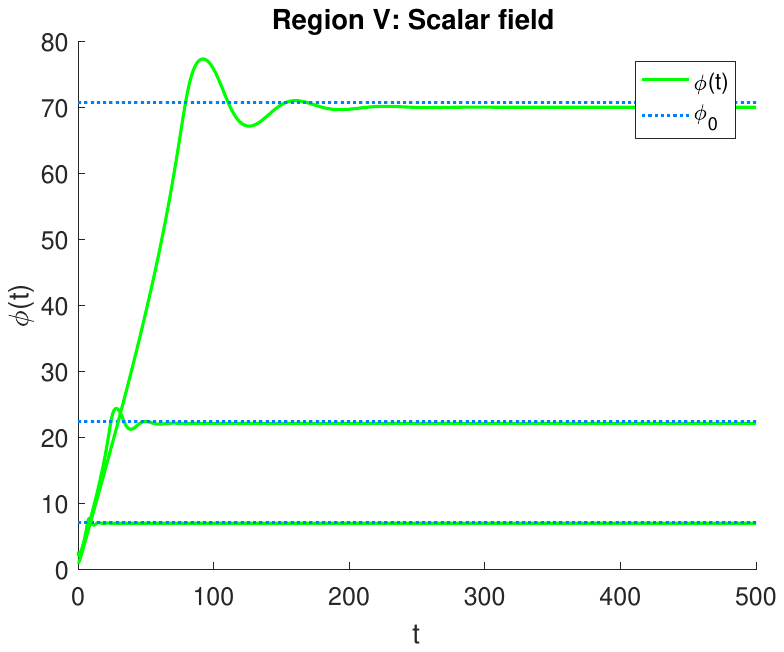}
		\caption{}
		\label{fig:Region5Scalar}
	\end{subfigure}%
\caption{Examples of dynamical solutions originally presented in \cite{Evnin}. Quantities are measured in \textit{bare} Planck units $(G=1)$, and the natural value for the vacuum energy, $G \Lambda=1$, has been assumed for concreteness. The solutions display a constant late-time behavior characterized by the constants $H_0\,, \phi_0$ given in \eqref{eq:region5} that are independent from the choice of initial conditions. Each green curve corresponds to a different value of the parameters satisfying \eqref{eq:parameters-conditions} and \eqref{eq:parameters-conditions-2}, and therefore asymptotes to different attractor values $H_0\,, \phi_0$ represented by the corresponding dashed blue line. (a) Dynamical behavior of the Hubble rate $H(t)$. (b) Dynamical behavior of the non-minimally coupled scalar field $\phi(t)$.}
\label{fig:Region5}
\end{figure}

It appears extremely hard, if not impossible, to come up with an EFT which would avoid large quantum contributions to the vacuum energy density, as in \eqref{eq:delta-rho}. However, one should keep in mind that only the Hubble expansion rate is actually measured, rather than the vacuum energy itself. In a previous publication \cite{Evnin}, we introduced an EFT which yields a \textit{dynamical relaxation} of the Hubble expansion rate from a Planckian value towards the observed small value. The model relies on a non-minimally coupled scalar field $\phi$, and is described by the Lagrangian\footnote{We use slightly different conventions than those of \cite{Evnin}, related by $16\pi G|_{\text{here}}=G|_{\text{there}}$ and $\lambda_R|_{\text{here}}=16\pi \lambda_R|_{\text{there}}$.}
\begin{equation}
\label{eq:action}
L=L_{EH}+L_{\phi}+L_m,
\end{equation}
where
\begin{gather}
    L_{EH} = \frac{1}{16\pi G} \left(R - 2\Lambda \right), \\
    \nonumber
    L_{\phi} = -\frac{1}{2} \left[(\partial_{\mu}\phi)^2 + m^2 \phi^2 + \xi R \phi^2 + \lambda \phi^4 + G\lambda_R R \phi^4\right],
\end{gather}
and $L_m$ accounts for conventional \text{minimally coupled} matter. It is characterized by the dimensionless parameters $G\Lambda, Gm^2, \xi, \lambda$ and $\lambda_R$, where $G\Lambda \sim 1$ is considered in order to avoid any fine-tuning problem associated with the cosmological constant. In addition, the model requires
\begin{equation}
\label{eq:parameters-conditions}
\xi <0 \qquad \text{and} \qquad 0\leq G m^2, \lambda, \lambda_R \ll |\xi|,
\end{equation}
where it is understood that $\lambda, \lambda_R$ cannot both vanish. When this is satisfied, \eqref{eq:action} was shown to possess de Sitter \textit{attractor solutions} characterized by a constant scalar field and Hubble expansion rate \cite{Evnin},
\begin{equation}
\label{eq:attractor}
H(t) \to H_0, \qquad \phi(t) \to \phi_0.
\end{equation}
Crucially, the value of the expansion rate $H_0$ is controlled by small ratios of parameters like $\lambda_R/|\xi|\ll 1$. This provides a cosmological scenario in which its small observed value is explained by small values of the coupling constants $m, \lambda, \lambda_R$, rather than by a small value of the cosmological constant $\Lambda$ itself as usually suggested by \eqref{eq:rho-observed}. 
This attractor behavior simply follows from the tendency of the scalar field to roll down towards the minimum of its potential. Indeed, although initially the spacetime curvature $R$ may be arbitrarily large (and possibly Planckian), the condition \eqref{eq:parameters-conditions} ensures a transient runaway behavior of $\phi$ accompanied by a decrease in potential energy, until the latter settles to its true minimum and becomes time-independent. Through Friedmann's equation, this decrease in potential energy implies a decrease in curvature, or equivalently, a decrease of the Hubble expansion rate $H(t)$. Equivalently, the scalar field $\phi$ adjusts its negative energy density in order to cancel most of the initially large vacuum energy, leaving only a small and positive remaining total energy density which is entirely controlled by the value of the coupling constants and accounts for the observed Hubble expansion. Figure~\ref{fig:Region5} reproduces some of the numerical solutions originally presented in \cite{Evnin}, displaying the approach to the constant attractor solutions. Finally, note that the action \eqref{eq:action} is a modification of the one put forward by Dolgov \cite{Dolgov:1982gh} and Ford \cite{Ford:1987de}, through the addition of the quartic couplings $\lambda, \lambda_R$. These are indeed crucial in order to avoid the unrealistic late-time vanishing of the gravitational interaction caused by an unbounded runaway behavior of $\phi$ present in the earlier model \cite{Dolgov:1982gh,Ford:1987de} and modifications thereof \cite{Dolgov:1982qq,Suen:1988nb}; we refer the reader to \cite{Evnin} for further details and comments on related work \cite{Barr:1986ya,Barr:2006mp,Dolgov:1996zg,Rubakov:1999bw,Emelyanov:2011wn,Emelyanov:2011kn,Emelyanov:2011ze,Dolgov:2008rf}.

\begin{figure}
\centering
	\begin{subfigure}[b]{.49\textwidth}
		\centering
		\includegraphics[scale=0.5]{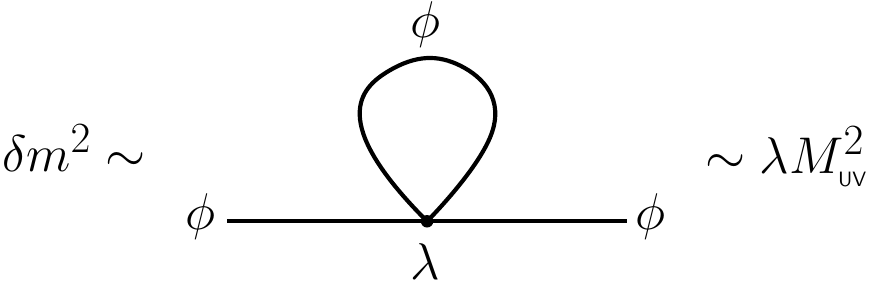}
		\vskip 3mm
		\caption{}
		\label{fig:Feynman1}
	\end{subfigure}
	\begin{subfigure}[b]{.49\textwidth}
		\centering
		\hskip 10mm
		\includegraphics[trim=0 0 1.1cm 0cm, scale=0.5]{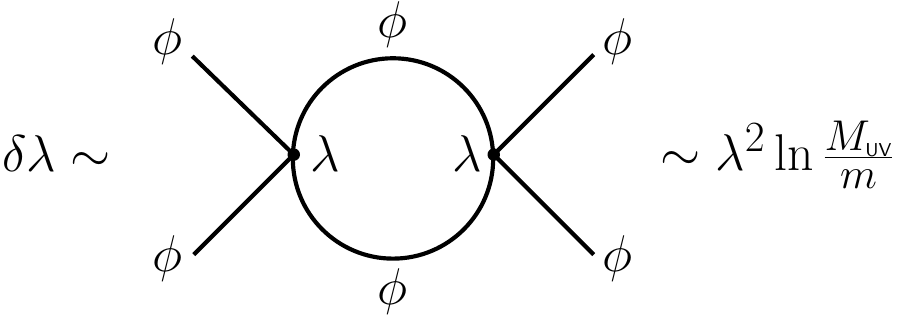}
		\caption{}
		\label{fig:Feynman2}
	\end{subfigure}%
\caption{One-loop Feynman diagrams giving corrections to (a) the scalar mass $m$ and (b) the quartic coupling $\lambda$.}
\label{fig:Feynman}
\end{figure}

The interest of the EFT \eqref{eq:action} is that it describes a small Hubble expansion rate while being \textit{technically} natural \cite{tHooft:1979rat}. Although it involves small dimensionless couplings, these do not require fine cancellations between various contributions in the UV, as one may assess by considering loop corrections to the small parameters $m, \lambda, \lambda_R$. In contrast to $\delta \rho_\Lambda$ in \eqref{eq:delta-rho}, these are proportional to the quartic couplings $\lambda$ and $\lambda_R$ themselves and are therefore as small (or smaller) than their tree-level values. Some of the corresponding one-loop Feynman diagrams are displayed in figure~\ref{fig:Feynman}. One may view the smallness of these quantum corrections as a pattern resulting from a \textit{weakly broken shift symmetry} $\phi \to \phi +c$. Indeed, quantum corrections to the symmetry-breaking terms $m, \lambda, \lambda_R$ must be proportional to interaction couplings that also break shift symmetry.\footnote{Note that $\xi$ does not need to be small since it only contributes to the propagator of the scalar field, and we will consider $|\xi| \sim 1$ for simplicity.}

In this paper, we extend our previous work and present a cosmological scenario based on the scalar-tensor theory \eqref{eq:action}, which features
\begin{itemize}
    \item an early cosmological era during which the attractor solution is reached,  
    \item a subsequent realistic cosmology through radiation- and matter-dominated eras,
    \item a gravitational interaction between massive bodies essentially indistinguishable from that of General Relativity,\footnote{Relaxation of the scalar-tensor theory \eqref{eq:action} towards a state indistinguishable from General Relativity, in the limiting case where $\Lambda=m=\lambda=0$, follows from an analysis performed by Damour and Nordtvedt \cite{Damour:1993id}. However, these authors did not consider the effect of a (large) cosmological constant, and the realization that a minimal modification of the same theory provides a resolution to the cosmological constant fine-tuning problem is new.}
    \item a small Hubble expansion rate at late-time as experimentally observed, without unnatural fine-tuning of parameters.
\end{itemize}
The dynamical approach towards the attractor solution during the early cosmological era is the one described in \cite{Evnin} and depicted in figure~\ref{fig:Region5}. In particular, any pre-existing minimally coupled matter completely dilutes by the time the attractor solution is reached. We assume a subsequent transition between this early attractor phase and radiation-domination, whose detailed description is beyond the scope of this work. One should expect it to include inflation and reheating as parts of it,\footnote{It would be very appealing if inflation could be described by the early attractor phase itself, in the model described here or a modification thereof.} upon which we further comment in the discussion. In practice, at this point of the cosmological history we simply insert the matter components appropriate to the description of matter-dominated eras, taking the attractor value $\phi_0$ as initial condition for the subsequent dynamical evolution of the scalar field (on the other hand the Hubble expansion rate is algebraically determined by the matter energy density through the modified Friedman equation). What we are left to show is that   
this dynamical evolution yields realistic cosmology and gravitational interaction between massive bodies, which is a priori highly non-trivial for a scalar-tensor theory such as the one described by \eqref{eq:action}. For this to be the case, it will be necessary to impose a refinement of the condition \eqref{eq:parameters-conditions}, namely
\begin{equation}
\label{eq:parameters-conditions-2}
\xi <0\,, \qquad \text{and} \qquad 0 \leq Gm^2, \lambda \ll \lambda_R \ll |\xi|\,,
\end{equation}
which was referred to as ``region 5'' of the parameter space in \cite{Evnin}. In this case, it was shown that the attractor values are given by
\begin{equation}
\label{eq:region5}
\phi_0^2=\frac{|\xi|}{2G \lambda_R}+O(Gm^2,\lambda)\,, \qquad H_0^2=\left(1+\frac{2\pi \xi^2}{\lambda_R} \right)^{-1}H_\Lambda^2 \ll H_\Lambda^2\,, 
\end{equation}
where $H_\Lambda^2\equiv \Lambda/3$ is the late-time Hubble expansion rate that one would obtain within General Relativity, but the present model yields the much lower value $H_0$. As mentioned before, we will consider $G\Lambda\,, |\xi| \sim 1$ for simplicity. A measurement of the Hubble expansion rate then amounts to a measurement of the non-minimal coupling $\lambda_R$, which will be shown to yield
\begin{equation}
\label{eq:lambdaR-10e60}
\lambda_R \sim 10^{-60}.
\end{equation}
With this parameter value, the model produces the observed Hubble expansion while accommodating for $G\Lambda \sim 1$, in stark contrast to the usual situation encountered within General Relativity. As previously discussed, such a small value is radiatively stable and is therefore \textit{technically} natural. One might wonder whether a more general class of models share the same attractive features and although this is very well possible, the present model is likely the simplest one of them. It is worth exploring how far one can go with this minimalistic scenario.

The paper is organized as follows. We start with an analysis of the evolution of the Hubble expansion rate $H(t)$ during matter eras in section~\ref{sec:cosmology}, assuming that the attractor value $\phi_0$ was (approximately) reached following an early attractor phase as described in \cite{Evnin}. In particular, we derive \eqref{eq:parameters-conditions-2} and \eqref{eq:lambdaR-10e60} as consistency conditions for a realistic cosmology, and show that $\phi(t)\approx \phi_0$ persists as stable solution of the equations of motion during matter eras. In section~\ref{sec:fifth force}, we consider the observational constraints on a `fifth force' mediated by the non-minimally coupled scalar field and show that these are automatically satisfied. In particular, we compute the post-Newtonian parameters $\gamma, \beta$ parametrizing the gravitational field of massive objects, as well as the effective gravitational strength measured between test masses, and show that they coincide with standard results obtained within General Relativity. We end with a discussion of the results in section~\ref{sec:discussion}, and point towards some open problems.

\section{\label{sec:cosmology} Cosmological evolution}
We assume spatial flatness, homogeneity and isotropy on cosmolgical scales, such that the system is described by a time-dependent scalar field $\phi(t)$, and a scale factor $a(t)$ of a Friedmann-Lemaitre-Robertson-Walker metric
\begin{equation}
ds^2=-dt^2+a(t)^2d\vec{x}^2, \qquad H(t)\equiv \frac{\dot{a}(t)}{a(t)}.
\end{equation}
The evolution of the scale factor is governed by\footnote{The modified Einstein equations also yield a second order differential equation for $a(t)$ that is automatically satisfied as it is identical to the conservation of the stress-energy tensor, a feature shared with General Relativity.}
\begin{equation}
\label{eq:Friedmann}
H^2 = \frac{8\pi G}{3}\left(\rho_\phi + \rho_m\right) + H_\Lambda^2, \qquad H_\Lambda^2\equiv \frac{\Lambda}{3},
\end{equation}
where $\rho_m$ is the energy density of conventional matter, and $\rho_\phi$ is the energy density of the non-minimally coupled scalar field, 
\begin{align}
    \rho_\phi= \frac{1}{2}&\big[(\partial_t \phi)^2+ m^2\phi^2+ \lambda\phi^4)+ (3H^2 + 3H\partial_t)(\xi\phi^2 + G\lambda_R \phi^4\big].
\end{align}
The equation of motion of the scalar field is given by
\begin{equation}
\label{eq:KG}
    \left[\partial_t^2 + 3H\partial_t + m^2 + 6\xi\left( \dot{H} + 2H^2 \right)\right]\phi 
    = -2\left[\lambda + 6G\lambda_R \left( \dot{H} + 2H^2 \right)\right]\phi^3.
\end{equation}
As usual, conventional matter includes various components, each of which has energy density $\rho_i$ and pressure $p_i$ that are related through an equation of state $p_i=w_i\, \rho_i$. The resulting total energy density takes the form
\begin{equation}
\label{eq:rho-m}
\rho_m(t)= \sum_i \rho_i(t)=\sum_i \rho_i(t_0)\, a(t)^{-n_i},
\end{equation}
with $n_i=3(1+w_i)$. In particular, non-relativistic matter and radiation obey $w=0$ and $w=1/3$, respectively.

We also assume that the attractor value $\phi_0$ in \eqref{eq:region5} has been reached long before the matter-dominated eras of interest, and determine the conditions under which $\phi(t)\approx \phi_0$ continues to be a stable solution of the system up until the present epoch of accelerated expansion. Below we find that these conditions automatically yield a cosmological evolution of the Hubble rate $H(t)$ essentially identical to that predicted by GR. For this we adopt the following parametrization of the scalar field,
\begin{equation}
\label{eq:delta-phi}
\phi(t) = \phi_0\left(1 - \delta \phi (t)\right),
\end{equation}
where $\delta \phi(t)$ will be understood as a dynamical perturbation of the attractor value $\phi_0$. We will show that such perturbations decay even after transition to radiation-domination, therefore ensuring stability of the proposed scenario.

As with any scalar-tensor theory, the effective Newton constant $\Geff$ differs from the constant $G$ appearing in the action \eqref{eq:action}, and is simply identified with the term multiplying the scalar curvature,
\begin{equation}
\label{eq:Geff-general}
\frac{1}{16\pi \Geff}=\frac{1}{16 \pi G}-\frac{1}{2}\left( \xi \phi_0^2+G \lambda_R \phi_0^4\right).
\end{equation}
Note that this simple identification only makes sense once the scalar field reaches a constant value $\phi_0$ such that the form of the Lagrangian \eqref{eq:action} indeed reduces to that of General Relativity. In particular, scalar fluctuations $\delta \phi(t)$ induce a nonzero time variation $\dot{G}_{\text{eff}}$ that is highly constrained. Using the appropriate generalization of \eqref{eq:Geff-general} to time-dependent situations, we prove in section~\ref{sec:fifth force} that such time variations stay well within observational bounds.
Although not immediately obvious, we also show that $\Geff$ is the physical Newton constant measured in small-scale experiments between massive bodies. In turn, one should keep in mind that physical quantities ought to be measured in physical Planck units. In particular, observation of the Hubble expansion rate amounts to
\begin{equation}
\Geff H_0^2 \sim 10^{-120},
\end{equation}
a seemingly unnaturally small value at the root of the cosmological constant problem. In the present model however, this will be accounted for by  natural values of the parameters.

The success of the $\Lambda$CDM model of cosmology suggests that the Hubble expansion rate should approximately satisfy Friedmann's equation sourced by conventional matter and a small `cosmological constant' parametrized by $H_0$,
\begin{equation}
\label{eq:delta-H}
H^2(t) \approx \frac{8\pi \Geff}{3} \rho_m(t)+H_0^2,
\end{equation}
where $\Geff$ and $H_0$ do not have to coincide with the constants $G$ and $H_\Lambda$ appearing in the action. This equation looks rather different from \eqref{eq:Friedmann} at first sight due to the presence of an additional energy component $\rho_\phi$, but we will demonstrate that they actually agree within the cosmological scenario proposed here. 

\subsection{Background attractor} Setting the linear scalar perturbation $\delta \phi$ to zero for the moment, \eqref{eq:KG} results in
\begin{equation}
\label{eq:phi0-full}
\phi_0^2 = -\frac{1}{2} \frac{6\xi(\dot{H}+2 H^2)+m^2}{6G\lambda_R (\dot{H}+2 H^2)+\lambda}.
\end{equation}
Since $H(t)$ is time-dependent, the only region in parameter space which is possibly consistent with the approximately constant solution $\phi(t)\approx \phi_0$ up to a small time-dependent deviation $\delta \phi(t)$, corresponds to
\begin{equation}
\label{eq:regionV}
\xi <0, \qquad G m^2,\lambda \ll \lambda_R \ll |\xi|.
\end{equation}
This was referred to as ``region 5'' of the parameter space in \cite{Evnin}. We will restrict our attention to this parameter space region from now on. The dimensionless couplings $Gm^2, \lambda$ will be used as expansion parameters characterizing the scalar perturbation $\delta \phi$. For simplicity, we use the generic notation $\varepsilon=\lbrace Gm^2, \lambda \rbrace$ as expansion parameter. Note also that $\varepsilon=0$ is a perfectly acceptable theory. At zeroth order, we therefore recover the attractor value \eqref{eq:region5} obtained in \cite{Evnin},
\begin{equation}
\label{eq:phi0}
\phi_0^2=\frac{|\xi|}{2G\lambda_R}.
\end{equation}
The attractor solution is unperturbed by the presence of matter density $\rho_m\neq 0$ at this order, since it is independent of $H(t)$. 

From equation \eqref{eq:Geff-general}, we find the relation between the bare and effective Newton's constants,
\begin{equation}
\label{eq:Geff-regionV}
\Geff=G\left(1+\frac{2\pi \xi^2}{\lambda_R} \right)^{-1}\ll G.
\end{equation}
Similarly, we deduce from \eqref{eq:Friedmann} that the late-time Hubble expansion rate, when $\rho_m \sim 0$, is
\begin{equation}
\label{eq:H0}
H(t) \to H_0\,, \quad H_0^2=\left(1+\frac{2\pi \xi^2}{\lambda_R} \right)^{-1}H_\Lambda^2 \ll H_\Lambda^2.
\end{equation}
When measured in effective Planck units, it yields
\begin{equation}
\label{eq:GeffH02}
\Geff H_0^2=\left(1+\frac{2\pi \xi^2}{\lambda_R} \right)^{-2}GH_\Lambda^2.
\end{equation}
Cosmological observations require this quantity to be of order $10^{-120}$. Considering for simplicity $G H_\Lambda^2, |\xi|\sim 1$ as already mentioned, we thus infer the value of the non-minimal coupling $\lambda_R$ to be
\begin{equation}
\lambda_R \sim 10^{-60}.
\end{equation}

\subsection{Friedmann's equation}
We turn to Friedmann's equation \eqref{eq:Friedmann} which we analyze up to first order in $\varepsilon=\lbrace Gm^2, \lambda \rbrace$. For this, we first compute the energy density of the scalar field,
\begin{equation}
\rho_\phi=\rho_{\phi_0}+\delta \rho_\phi+O(\varepsilon^2),
\end{equation}
where
\begin{subequations}
\begin{align}
\rho_{\phi_0}&=3H^2\left(\xi \phi_0^2+G\lambda_R \phi_0^4\right)=-\frac{3\xi^2}{4G\lambda_R} H^2,\\
\delta \rho_\phi&=\phi_0^2 \left[\frac{1}{2}\left(m^2+\lambda \phi_0^2 \right)-\left(2\xi+4G\lambda_R \phi_0^2 \right)\left(3H^2+3H \partial_t\right) \delta \phi(t) \right]\\
&=\frac{1}{2G} \left(Gm^2+\frac{|\xi| \lambda}{2\lambda_R} \right) \phi_0^2.
\end{align}
\end{subequations}
Note that the term linear in $\delta \phi(t)$ explicitly disappears thanks to \eqref{eq:phi0}. Plugging this into Friedmann's equation \eqref{eq:Friedmann}, we find
\begin{align}
\label{eq:Friedmann-2}
H^2&=\frac{8\pi \Geff}{3}\left(\rho_m+\delta \rho_\phi\right)+H_0^2+O(\varepsilon^2)=\frac{8\pi \Geff}{3}\rho_m+\Heff^2+O(\varepsilon^2).
\end{align}
This takes the form of the standard Friedmann's equation with \textit{renormalized} Newton constant $\Geff$, and effective `cosmological constant' term $\Heff$ given by
\begin{align}
\Heff^2\equiv H_0^2\left(1+\delta H_0^2\right),
\end{align}
with
\begin{equation}
\delta H_0^2=\frac{8\pi \Geff}{3H_0^2}\delta \rho_\phi=\frac{2\pi |\xi|}{3\lambda_R GH_\Lambda^2}\left(Gm^2+\frac{|\xi|\lambda}{2\lambda_R}\right).
\end{equation}
The term $\delta H_0^2$ is an order $O(\varepsilon)$ correction to the leading value of the late-time Hubble expansion rate $H_0$ described in \eqref{eq:H0}. It should be subleading for the perturbative expansion to make sense, which requires
\begin{equation}
\label{eq:condition-lambda}
\lambda \ll \lambda_R^2 \sim 10^{-120}.
\end{equation}
This condition further restricts the allowed parameter space.

In summary, given that the attractor value $\phi_0$ was reached prior to the cosmological eras of interest, the evolution of the scale factor in the theory \eqref{eq:action} satisfies the standard equation of Friedmann sourced by conventional matter. However, the effective parameters $\Geff$ and $\Heff$ entering the latter do not coincide with the bare parameters $G$ and $H_\Lambda$ appearing in the action. In particular, $\Geff \Heff^2 \sim 10^{-120}$ even though $G H_\Lambda^2 \sim 1$.

\subsection{Linear perturbations}
In order to assess the stability of the approximately constant solution $\phi(t) \approx \phi_0$, we have to make sure that scalar perturbations $\delta \phi(t)$ stay small at all times. These are governed by the evolution equation 
\begin{equation}
\label{eq:deltaphi-evolution}
    \left[\partial_t^2 + 3H\partial_t + 12|\xi|\left(\dot{H}+2H^2\right) \right]\delta\phi = m^2+\frac{|\xi|\lambda}{G\lambda_R},
\end{equation}
obtained at linear order in $\epsilon$ from \eqref{eq:KG}.
To analyze this equation, it is useful to make a change of time variable\footnote{\label{footnote-2} This change of variable is ill-defined if $\dot{H}+2H^2=0$. A purely radiation-dominated era  with $H=1/2t$ is a particular case thereof, for which the general solution to \eqref{eq:deltaphi-evolution} is given by
\begin{equation*}
\delta \phi(t)=\frac{Gm^2+|\xi|\lambda/\lambda_R}{5G}\, t^2+C_1\, t^{-1/2}+C_0\,,
\end{equation*}
where $C_0, C_1$ are integration constants. One can argue that a Universe filled with only radiation never occurred, such that the above discussion is purely academic.}
\begin{equation}
\tau=\int_0^t dt'\ w(t'), \qquad w(t)=\sqrt{12|\xi|\left(\dot{H}+2H^2\right)}, 
\end{equation}
such that \eqref{eq:deltaphi-evolution} becomes
\begin{equation}
\label{eq:deltaphi-evolution-2}
\left[\partial_\tau^2 +f(\tau)  \partial_\tau + 1\right]\delta\phi(\tau) =\delta \phi_p(\tau),
\end{equation}
with
\begin{equation}
\label{eq:friction}
f\equiv 6|\xi|w^{-3}\left(\ddot{H}+10H \dot{H}+12H^3\right), \qquad \delta \phi_p\equiv \frac{Gm^2+|\xi|\lambda/\lambda_R}{12 |\xi|G(\dot{H}+2H^2)}.
\end{equation}
This equation is that of a damped harmonic oscillator with time-dependent friction $f(\tau)$ and source $\delta \phi_p(\tau)$. We will show that $\delta \phi$ tends towards $\delta \phi_p$ as time evolves,
\begin{equation}
\label{eq:deltaphi-p}
\delta \phi(t) \quad \longrightarrow \quad \delta \phi_p(t).
\end{equation}
This linear correction $\delta \phi_p$ is identified with the order $O(\varepsilon)$ time-dependent correction to the constant attractor value $\phi_0$, consistently with the expression \eqref{eq:phi0-full}, and decreases to a small constant in the late-time limit $H(t) \to H_0$. Showing the convergence \eqref{eq:deltaphi-p} will therefore ensure stability of the attractor solution, and will actually directly follow from positivity of the friction term $f(t)$ in \eqref{eq:deltaphi-evolution-2}. To show this, we first introduce the density parameters
\begin{equation}
\Omega_i(t)\equiv \frac{8\pi \Geff}{3H(t)^2}\, \rho_i(t), \qquad \Omega_\Lambda(t) \equiv \frac{H_0^2}{H(t)^2},
\end{equation}
and we note that Friedmann's equation \eqref{eq:Friedmann-2} implies
\begin{equation}
\sum_i \Omega_i(t)+\Omega_\Lambda(t)=1+O(\varepsilon).
\end{equation}
Using \eqref{eq:rho-m}, we can show that the friction term  \eqref{eq:friction} may be rewritten
\begin{equation}
f(t)=\frac{1}{8\sqrt{24 |\xi|}} \left(1-\frac{1}{4} \sum_i n_i \Omega_i(t)\right)^{-\frac{3}{2}} \left[24\,  \Omega_\Lambda(t)+\sum_i \left(n_i^2-10 n_i+24\right)\Omega_i(t)\right]+O(\varepsilon).
\end{equation}
Each term in the sum within bracket is strictly positive if $n_i < 4$ for each matter constituent, such that $f(t)>0$ at all times. This is equivalent to the condition on the equation of state $w_i < 1/3$, satisfied by conventional matter (the limiting case of pure radiation $w=1/3$ is described in footnote~\ref{footnote-2}).

Therefore, the `energy'
\begin{equation}
E\equiv\frac{1}{2}\left[ (\delta \phi-\delta \phi_p)^2+(\partial_\tau \delta \phi)^2\right],
\end{equation}
of this harmonic oscillator satisfies the evolution equation
\begin{equation}
\label{eq:dE}
\partial_\tau E=-f(\tau) (\partial_\tau \delta \phi)^2-(\delta \phi-\delta \phi_p) \partial_\tau \delta \phi_p.
\end{equation}
If this quantity was negative at all times, then the convergence $\delta \phi \to \delta \phi_p$ would directly follow. Indeed, one has 
\begin{equation}
\left(\delta \phi -\delta \phi_p\right)^2 < E,
\end{equation}
such that a continuous decrease in energy $E$ implies that $|\delta \phi -\delta \phi_p|$ is bounded from above by an ever decreasing function asymptoting to zero. For $m=\lambda=0$, one has $\delta \phi_p=0$ such that only the first term on the right-hand side of \eqref{eq:dE} is present. Since the friction $f(\tau)$ is strictly positive for conventional matter, $\partial_\tau E<0$ at all times and any scalar perturbation $\delta \phi$ eventually decays to zero. For $m, \lambda \neq 0$, the second term on the right-hand side of \eqref{eq:dE} can be either positive or negative. Although this might lead to very brief periods of energy increase when $\partial_\tau \delta \phi \sim 0$, its effect should still be negligible on average as it is of order $O(\varepsilon)$. We therefore expect $\partial_\tau E <0$ and $\delta \phi \to \delta \phi_p$ in general.

In summary, small scalar perturbations generated during matter eras do not grow to eventually destabilize the constant attractor solution $\phi_0$, provided that $\delta \phi_p \ll 1$. This requires 
\begin{equation}
\label{eq:additional-condition-1}
\max \Big\lbrace Gm^2, |\xi|\lambda/\lambda_R \Big\rbrace \ll |\xi| G \left( \dot{H}+2H^2\right),
\end{equation}
or equivalently,
\begin{equation}
\label{eq:additional-condition}
\max \Big\lbrace Gm^2 \lambda_R, |\xi|\lambda \Big\rbrace \ll |\xi|^3\Geff \left( \dot{H}+2H^2\right).
\end{equation}
This additional condition is easily satisfied by small enough $Gm^2$ and $\lambda$. At late times in particular, $H \to H_0$ and using \eqref{eq:GeffH02}, one can show that it simply reduces to the previous condition \eqref{eq:condition-lambda}.

\section{Fifth force constraint \label{sec:fifth force}}
We now turn to the study of gravitational effects associated with the scalar-tensor theory \eqref{eq:action} that could be observed on small scales through solar system experiments, for example. Since there is no direct coupling between the scalar field $\phi$ and the matter sector, massive bodies made of conventional matter still satisfy the equivalence principle. However, the presence of the non-minimally coupled scalar field may alter the gravitational field sourced by any such massive body compared to the one predicted by General Relativity. To characterize such deviations, we will use the parametrized post-Newtonian (PPN) formalism, whose long history starts with the classic work of Eddington \cite{Eddington}; see the textbooks \cite{PoissonWill,Will:2018bme} and references therein. This formalism has been applied to general scalar-tensor theories of gravity in \cite{Damour:1992we}. Hence, we simply have to specialize the general results obtained in that work to the model at hand. We will follow the conventions and formulas of \cite{Boisseau:2000pr}. Gravitational experiments performed within the solar system result in constraints on the PPN parameters, and therefore in constraints on the ``fifth force'' mediated by the scalar degree of freedom.

The Lagrangian density \eqref{eq:action} may be written as
\begin{equation}
\label{eq:Lagrangian}
L = \frac{1}{2} \left[F(\phi) R - \left(\partial_\mu \phi\right)^2 \right] - U(\phi) + L_m,
\end{equation}
with
\begin{align}
    F(\phi) &= \frac{1}{8\pi G}-\left(\xi \phi^2 + G\lambda_R \phi^4\right),\\
    U(\phi) &=\frac{\Lambda}{8\pi G}+\frac{1}{2} \left(m^2 \phi^2 + \lambda \phi^4\right).
\end{align}
We restrict the discussion to parameters satisfying the conditions \eqref{eq:regionV}, \eqref{eq:condition-lambda} and \eqref{eq:additional-condition}, since these yield a realistic cosmological evolution from radiation-dominated eras onward.  

\subsection{Gravitational coupling between test masses} 
The first important comment is that the effective gravitational coupling $\Geff'$ between two test masses that does not necessarily coincide with the effective Newton constant $\Geff$ identified in \eqref{eq:Geff-regionV} for the parameter space of interest, a common issue shared by scalar-tensor theories. Instead, it is given by \cite{Boisseau:2000pr}
\begin{equation}
\label{eq:Geff-prime}
G'_{\text{eff}}= \frac{1}{8\pi F} \left( \frac{2F + 4 (dF/d\phi)^2}{2F + 3(dF/d\phi)^2}\right),
\end{equation}
an expression valid over length scales $l$ satisfying
\begin{align}
l^{-2} H^{-2} \gg \max \Big\lbrace 1,\left|\frac{d^2 F}{d\phi^2}\right|,H^{-2}\left|\frac{d^2U}{d\phi^2} \right| \Big\rbrace\bigg|_{\phi_0}\approx \max \Big\lbrace 1,4|\xi|,\frac{Gm^2+3|\xi|\lambda/\lambda_R}{GH^2} \Big\rbrace.  
\end{align}
Hence, \eqref{eq:Geff-prime} holds on sub-Hubble scales. We want to evaluate it for the cosmological scenario at hand. For this, we first compute the field derivatives of $F$ evaluated for the cosmological solution \eqref{eq:delta-phi} at linear order in the scalar perturbation,
\begin{align}
F&=\frac{1}{8\pi \Geff}+O(\delta \phi^2),\\
\frac{dF}{d\phi}&=4|\xi|\phi_0\, \delta \phi+O(\delta \phi^2).
\end{align}
Plugging this into \eqref{eq:Geff-prime}, we find
\begin{equation}
\label{eq:Geff-agreement}
\Geff'=\Geff+O(\delta \phi^2).
\end{equation}
Hence, the two effective gravitational couplings essentially coincide within the considered cosmological scenario.

It also follows from \eqref{eq:Geff-agreement} that $\Geff'$ is constant up to linear order $\delta \phi$. This is another desirable property since no time variation of Newton's constant has been ever observed so far. The most stringent bounds come from measurements of Mars ephemeris, yielding $\dot{G}'_{\text{eff}}/\Geff'=0.1\pm 1.6 10^{-13}\, \text{yr}^{-1}$ \cite{Konopliv:2011} and $-0.6\pm0.4 10^{-13}\, \text{yr}^{-1}$ \cite{Pitjeva:2013}. See \cite{Will:2018bme} for an exhaustive review of various observational tests. Constraints on the parameters of the model can in principle be obtained by going to quadratic order in perturbation theory, but we do not pursue this here.

\subsection{Post-Newtonian parameters}
The the first post-Newtonian corrections to the gravitational field of a massive source are parametrized by $\gamma$ and $\beta$, which encode the leading deviations from the Schwarzschild metric in a $1/r$ radial expansion. In the context of scalar-tensor theories, they are given by \cite{Boisseau:2000pr}
 \begin{align}
 \gamma&=1- \frac{ (dF/d\phi)^2}{2F + 2(dF/d\phi)^2},\\
\beta &=1+\frac{1}{4} \frac{F (dF/d\phi)}{2F + 3(dF/d\phi)^2} \frac{d\gamma}{d \phi}.
 \end{align}
For the cosmological solution of interest and up to linear order in scalar perturbations, their values simply coincide with those of General Relativity, 
\begin{subequations}
\label{eq:gamma-beta}
 \begin{align}
 \gamma &=1+O(\delta \phi^2),\\
\beta&=1+O(\delta \phi^2).
 \end{align}
\end{subequations}

The best experimental constraint on post-Newtonian parameters comes from Doppler tracking of the Cassini spacecraft \cite{Bertotti:2003rm,Will:2018bme},
\begin{equation}
\label{eq:Cassini}
\gamma -1=(2.1\pm 2.3)\times 10^{-5}.
\end{equation}
This constraint, together with a careful analysis of the quadratic corrections to \eqref{eq:gamma-beta}, would put a precise bound on the amplitude of scalar perturbations $\delta \phi$ in the solar sytem. Roughly speaking, these quadratic corrections are of order
\begin{equation}
\gamma-1 \approx \Geff\, \phi_0^2\, \delta \phi^2 \approx \delta \phi^2.
\end{equation}
We recall that $\delta \phi(t)$ tends to $\delta \phi_p(t) \ll 1$ described in \eqref{eq:deltaphi-p}, which does not yield a violation of the Cassini bound \eqref{eq:Cassini} provided that $Gm^2, \lambda$ are sufficiently small. On the other hand, there is no guarantee that initially larger scalar perturbations cannot be generated at the present epoch. This offers some prospects of actually testing the model under consideration, which is otherwise indistinguishable from General Relativity. 

\section{Discussion \label{sec:discussion}}

We have described a cosmological scenario based on the scalar-tensor theory \eqref{eq:action}, in which the attractor value of the scalar field $\phi_0$ is assumed to be (approximately) reached long before radiation-dominated eras. Within the appropriate parameter regime \eqref{eq:parameters-conditions-2}, we have shown that the subsequent evolution of the scale factor satisfies Friedmann's equation sourced by conventional matter, and that post-Newtonian parameters characterizing the gravitational interaction on small scales are essentially identical to those of General Relativity. Importantly though, the late-time Hubble expansion rate $H_0$ can take the small observed value $\Geff H_0^2 \sim 10^{-120}$ even though the bare cosmological constant is Planckian (in bare units), i.e., $G \Lambda \sim 1$. In this way, it provides a potential resolution to the cosmological constant fine-tuning problem.

We have shown that linear scalar perturbations generically decay over the cosmological history. Identification of potential sources of scalar perturbations would offer the opportunity to further test the proposed model. In particular, measurements of the PPN parameter $\gamma$ are sensitive to quadratic corrections $\delta \phi^2$, a fact which might be used to derive additional constraints in case that sizeable perturbations are being produced at the current epoch.  

We have focused our attention to cosmological evolution for spatially homogeneous fields as a first approximation. Of course, it will be interesting to include inhomegeneities in view of their crucial relevance to large scale structure formation. In this regard, we anticipate that standard
cosmology derived within General Relativity may again be recovered (as
we have shown for spatially uniform solutions), provided that the
damping phase has already occurred and that the attractor value $\phi_0$ has been
reached with high precision. Under this assumption, all we need to check is that this constant attractor value $\phi_0$ continues to be a valid solution of the equations of motion even in the presence of inhomogeneities in the conventional matter sector and the metric. The general form of the Klein--Gordon equation \eqref{eq:KG} for arbitrary metric configurations is 
\begin{equation}
\label{eq:KG-general}
    \left(\square - m^2 - \xi R\right)\phi 
    = 2\left(\lambda + G\lambda_R R\right)\phi^3.
\end{equation}
Setting $m=\lambda=0$ and inserting $\phi(t,x)=\phi_0$ given in \eqref{eq:phi0}, this equation is indeed satisfied for \textit{any} metric configuration. At least for this ideal choice of parameters, we conclude that conventional cosmology may be embedded within the proposed scalar-tensor theory. Small deviations from standard cosmology may arise as a result of $O(G m^2,\lambda)$ corrections or sources of inhomogeneous scalar fluctutations $\delta \phi(t,x)$ which would need to be identified. It will be interesting to study whether such small corrections stay within observational bounds and whether they could actually improve the $\Lambda$CDM cosmological model, for instance by resolving tensions between independent measurements of the Hubble expansion rate \cite{Riess:2019cxk,Verde:2019ivm,Wong:2019kwg}.

From the EFT perspective, naturalness of the scalar-tensor theory \eqref{eq:action} holds since loop corrections to the small parameters $m, \lambda, \lambda_R$ are suppressed by powers of $\lambda$ and $\lambda_R$ themselves. However, we have treated the gravitational field as purely classical and we have not considered the potential effect of quantum gravitational fluctuations. It is an interesting open question to assess whether one-loop gravitational corrections could generate new interactions terms - such as matter-scalar couplings - that would question the naturalness of the scalar-tensor theory under consideration. Note that if the vacuum energy parametrized by $\Lambda$ is not Planckian (in bare units) as we have assumed throughout this paper, but at a significantly smaller scale like that of a Grand Unification for example, the proposed damping mechanism of the Hubble rate would still apply without being threatened by quantum gravitational corrections.

We view the present results as a strong case for further investigations of the model or modifications thereof, including more detailed comparisons against available data originating from various epochs of the Universe's history. This would include big bang nucleosynthesis as the present analysis does not enable us to draw any conclusion in that respect. Another open problem is to connect the proposed cosmological scenario to the theory of inflation.\footnote{The scalar-tensor theory \eqref{eq:action} restricted to $\lambda=\lambda_R=\Lambda=0$, in conjunction with an inflaton field, has been recently considered in \cite{Ringeval:2019bob} as a way to generate the large observed hierarchy between the measured Planck mass and Hubble expansion in a way analogous to \eqref{eq:GeffH02}. In contrast to the present work, a discussion of the cosmological constant fine-tuning problem is missing since this would require to consider a large (Planckian) $\Lambda$. It would be interesting to assess whether a construction along the same lines could incorporate the resolution of the cosmological constant fine-tuning problem proposed here.} In particular, one should ensure that the inflationary paradigm is not in tension with some assumptions made here. We hope to get back to this important question in the future.

\section*{Acknowledgments}
We thank Jos\'e Espinosa, Miguel Montero and Thomas Van Riet for discussions on related topics. This work was supported in part by FWO-Vlaanderen through project G006918N. The work of OE is funded by CUniverse research promotion project (CUAASC) at Chulalongkorn University. The work of VM is supported by NSERC of Canada and by the Department of Physics and the Faculty of Graduate Studies of the Universit\'e de Montr\'eal. The work of KN is supported by a grant from the John Templeton Foundation and a by grant from the Science and Technology Facilities Council (STFC).

\bibliography{CCDampingScenario}
\bibliographystyle{utphys.bst}
\end{document}